\documentclass[12pt, prd, showpacs]{revtex4}
\usepackage{amssymb}
\usepackage{graphicx}
\usepackage{dcolumn}
\usepackage{bm}
\usepackage{color}
\usepackage{amsmath}
\usepackage{array}

\begin{document}

\title{Super-Penrose process for extremal rotating neutral white holes}
\author{O. B. Zaslavskii}
\affiliation{Department of Physics and Technology, Kharkov V.N. Karazin National
University, 4 Svoboda Square, Kharkov 61022, Ukraine}
\affiliation{Institute of Mathematics and Mechanics, Kazan Federal University, 18
Kremlyovskaya St., Kazan 420008, Russia}
\email{zaslav@ukr.net }

\begin{abstract}
We consider collision of two particles 1 and 2 near the horizon of the
extremal rotating axially symmetric neutral generic black hole producing
particles 3 and 4. We discuss the scenario in which both particles 3 and 4
fall into a black hole and move in a white hole region. If particle 1 is
fine-tuned, the energy $E_{c.m.}$ in the centre of mass grows unbounded (the
Ba\~{n}ados-Silk-West effect). Then, particle 3 \ can, in principle, reach a
flat infinity in another universe. If not only $E_{c.m.}$ but also the
corresponding Killing energy $E$ is unbounded, this gives a so-called
super-Penrose process (SPP). We show that the SPP\ is indeed possible. Thus
white holes turn out to be potential sources of high energy fluxes that
transfers from one universe to another. This generalizes recent observaitons
made by Patil and Harada for the Kerr metric. We analyze two different
regimes of the process on different scales.
\end{abstract}

\keywords{particle collision, super-Penrose process, white holes}
\pacs{04.70.Bw, 97.60.Lf }
\maketitle

\section{Introduction}

During last decade a lot of work has been made for investigation of
properties of high energy processes near black holes. This was stimulated by
the paper \cite{ban}, where it was found that collision of two particles
near rapidly rotating black holes can lead to unbounded energies $E_{c.m.}$
in the centre of mass frame. This was called the Ba\~{n}ados-Silk-West (BSW)
effect, after the authors' names. After its publication, it turned out, that
there are also earlier works \cite{katz}, \cite{ps} in which near-horizon
particle collisions in the Kerr metric were investigated. Meanwhile, a
typical process considered there, includes head-on collision between two
arbitrary particles, where particle 1 arrives from infinity while particle 2
comes from the horizon (see eq. 2.57 of \cite{ps}). But as far as particle 2
is concerned, this is nothing else than a typical behavior of a particle
near a white hole. If such a region is allowed in the complete space-time,
the effect of unbounded $E_{c.m.}$ for head-on collisions exists \ even in
the Schwarzschild metric \cite{tot}, \cite{white}. Thus white holes can be
an alternative to black ones as a source of high energy collisions.

More important question is whether it is possible to gain not only unbounded 
$E_{c.m.}$ but also unbounded conserved Killing energies $E$ since it is the
latter quantity which can be measured in the Earth laboratory, at least in
principle. The collisions in the Schwarzschild background are useless for
this purpose since energy cannot be extracted at all. For such an
extraction, the existence of negative energies and ergoregion are required
that makes it possible the Penrose process \cite{pen} or its collisional
analogue \cite{col}. If the energy gain is unbounded, this is called the
super-Penrose process (SPP). For black holes, the energy gain is finite, so
the SPP is impossible for them (see \cite{is} and references therein).

In this context, there is a scenario with participation of white holes,
different from those in \cite{katz} - \cite{white}. Now, both particles
collide near the black hole horizon in "our" part of Universe but afterwards
the products of reaction leave it. Passing though the horizon, they appear
inside a white whole region and, eventually, transfer energy to another
universe. Or, vice verse, collision in another universe can give rise to
high energy in our one. If such a process is possible, this would give
astrophysical realization of high energy transfer with white holes as a
source that was suggested earlier \cite{nov}, \cite{dad}.

The concrete process of this kind in the Kerr background was considered
recently in \cite{ph}. The authors showed that the conserved energy of
produced particles can be as large as one like. In other words, the SPP is
possible. Our aim is to extend consideration to generic rotating axially
symmetric stationary white holes. In doing so, we exploit the approach that,
in our view, is simpler and was already used for examination of the energy
extraction from generic black holes of the aforementioned type including the
Kerr metric \cite{up}, \cite{en}. In particular, we do not use
transformation between the three frames (center of mass, locally
non-rotating and stationary ones) and work in the original frame.

Although there are reasons to believe that white holes are unstable \cite%
{unst} (see also Sec. 15 of \cite{fn}), motivation for consideration of such
objects stems from different roots. (i) High energy process, if they are
confirmed, can themselves contribute to the instability of white holes, so
they are important for elucidation of the fate of such objects. (ii) The
complete theory of the BSW effect should take into consideration all
possible configurations and scenarios, at least for better understanding the
phenomenon.

We use the system of units in which the fundamental constants $G=c=1$.

\section{Basic equations}

Let us consider the metric%
\begin{equation}
ds^{2}=-N^{2}dt^{2}+\frac{dr^{2}}{A}+g_{\phi }(d\phi -\omega
dt)^{2}+g_{\theta }d\theta ^{2}\text{,}  \label{met}
\end{equation}%
where $g_{\phi }\equiv g_{\phi \phi }$, $g_{\theta }\equiv g_{\phi \phi }$,
all coefficients do not depend on $t$ and $\phi $. Before turning to the
analysis of the collisions between two particles in the background of a
white hole, we describe main features of motion inherent to an individual
particle.

To simplify formulas, we assume that in the equatorial plane $A=N^{2}$.
Otherwise, we can always achieve this equality by redefining the radial
coordinate according to $r\rightarrow \tilde{r}$, where%
\begin{equation}
\frac{dr}{\sqrt{A}}=\frac{d\tilde{r}}{N}\text{,}
\end{equation}%
so%
\begin{equation}
\tilde{r}=\int^{r}\frac{dr^{\prime }N}{\sqrt{A^{\prime }}}\text{.}
\end{equation}

In general, for an arbitrary $\theta $, this transformation does not work
since $d\tilde{r}$ would not be a total differential, if $A$ and $N$ depend
on $\theta $. However, for our purposes (for motion within the equatorial
plane, so $\theta =\frac{\pi }{2}$ is fixed), this is a quite legitimate
operation. It is valid for any metric of the type (\ref{met}) including the
Kerr one.

For a given energy $E$, angular momentum $L$ and mass $m$ the equation of
motion in the equatorial plane read%
\begin{equation}
m\dot{t}=\frac{X}{N^{2}}\text{,}
\end{equation}%
\begin{equation}
X=E-\omega L\text{,}  \label{X}
\end{equation}%
\begin{equation}
m\dot{\phi}=\frac{L}{g_{\phi }}
\end{equation}%
\begin{equation}
m\dot{r}=\sigma P\text{, }\sigma =\pm 1\text{,}
\end{equation}%
\begin{equation}
P=\sqrt{X^{2}-\tilde{m}^{2}N^{2}}\text{,}  \label{P}
\end{equation}%
\begin{equation}
\tilde{m}^{2}=m^{2}+\frac{L^{2}}{g_{\phi }}\text{,}
\end{equation}%
dot denotes differentiation with respect to the proper time $\tau $. TThe
forward-in-time condition requires%
\begin{equation}
X\geq 0\text{.}  \label{ftc}
\end{equation}

In what follows, we will use the standard classification of particles. If $%
X_{H}=0$, a particle is called critical. If $X_{H}=O(1)$, it is called
usual. If $X_{H}=O(N_{c}),$ it is called near-critical. Here, subscripts "H"
and "c" refer to the quantities calculated on the horizon and the point of
collision, respectively.

For the near-critical particle, we use presentation%
\begin{equation}
L=\frac{E}{\omega _{H}}(1+\delta )  \label{LE}
\end{equation}%
exploited in \cite{en}. Here,%
\begin{equation}
\delta =C_{1}N_{c}  \label{del}
\end{equation}%
is a small quantity for collisions near the horizon, $C_{1}=O(1)$ is a
constant.

Near the horizon, we assume the Taylor expansion that for the extremal case
reads \cite{tan}%
\begin{equation}
\omega =\omega _{H}-B_{1}N+O(N^{2})\text{.}
\end{equation}

Then, we have the following approximate expressions there.

The critical particle: 
\begin{equation}
X=\frac{b}{h}EN+O(N^{2})\text{,}  \label{cr}
\end{equation}%
\begin{equation}
P=N\sqrt{E^{2}\left( \frac{b^{2}-1}{h^{2}}\right) -\frac{1}{h^{2}}-m^{2}}.
\end{equation}

A\ usual particle:%
\begin{equation}
X=X_{H}+B_{1}LN+O(N^{2})\text{, }
\end{equation}%
\begin{equation}
X_{H}=E-\omega _{H}L\text{,}
\end{equation}%
\begin{equation}
P=X+O(N^{2}).  \label{us}
\end{equation}

The near-critical particle:%
\begin{equation}
X=E(\frac{b}{h}-C_{1})N  \label{xncr}
\end{equation}%
\begin{equation}
P=N\sqrt{E^{2}[(\frac{b}{h}-C_{1})^{2}-\frac{1}{h^{2}}]-m^{2}}+O(N^{2})\text{%
.}  \label{ncr}
\end{equation}

We introduced notations $b=B_{1}\sqrt{g_{H}}$, $h=\omega _{H}\sqrt{g_{H}}$.
Here, for shortness, we also use notation $g\equiv g_{\phi }$, where $%
g_{\phi }$ is defined in (\ref{met}).

To give an example, we list the concrete expressions for these relevant
characteristic for the physically relevant case of the extremal Kerr-Newman
metric:%
\begin{equation}
g_{H}=\frac{(M^{2}+a^{2})^{2}}{M^{2}}\text{,}
\end{equation}%
\begin{equation}
B_{1}=\frac{2a}{M^{2}+a^{2}}\text{,}
\end{equation}%
\begin{equation}
\omega _{H}=\frac{a}{M^{2}+a^{2}}\text{,}
\end{equation}%
\begin{equation}
b=\frac{2a}{M}\text{, }h=\frac{a}{M}\text{.}  \label{bh}
\end{equation}

Here, $a$ is the standard parameter of the Kerr-Newman metric that
characterizes its angular momentum, $M$ being the mass. It is interesting
that the electric charge $Q$ drops out from the explicit expressions for $b$
and $h$. If $Q=0$, we our extremal metric transforms to the Kerr one with $%
M=a$ and the values (\ref{bh}) return to $b=2$ and $h=1$ in accordance with
eq. (46) of \cite{en}.

Meanwhile, below we operate with relevant quantities in a general form,
without specifying the metric. This is justified by the fact that, as we
will see, in the cases under discussion there is no concrete upper bound on
the maximum possible energy, this result being model-independent.

It is worth noting that the coordinates in (\ref{met}) generalize the
Boyer-Lindquiste ones for the Kerr metric. It is known that such coordinates
do not cover the whole space-time. Therefore, one is led to introduce an
infinite set of different coordinate patches, the space-time structure
includes an infinite set of black hole and white hole regions. For the, say,
extremal Kerr-Newman metric for the equatorial plane the Carter-Penrose
diagrame is similar to that for the Reissner-Nordstr\"{o}m metric and is
represented on Fig. 1 (see, e.g. detailed description of these metrics in 
\cite{exact}, especially Ch. 11). We schematically showed a trajectory of a
test particle \textbf{Fig. 1}.
\begin{figure}
\begin{center}
\includegraphics[width=0.5\textwidth]{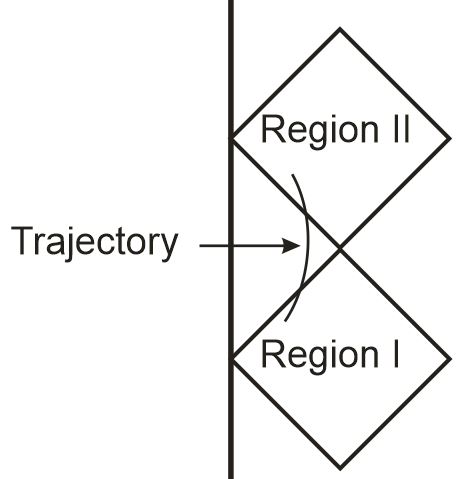}
\caption{The Carter-Penrose diagrame of the extremal Kerr-Newman space-time in the equatorial plane.}
\label{penrose}
\end{center}
\end{figure}
The concrete descreiption of motion inside the horizon was done in \cite{ph}
with the help of ligh-like coordinates for the Kerr metric. Although such details are
of interest on their own right, for our goal (to elucidate the absence or
prsence of the upper bound on $E$) they are not necessary, so we use more
simple and straightforward approach. It is based on already elaborated
scheme exploited for the analysis of collisions in the black hole background 
\cite{en}, \cite{rn}. 

\section{Scenario of collision}

In this section, we give brief set-up for the description of the process
under consideration. We assume that particles 1 and 2 collide producing
particles 3 and 4. We want to elucidate, whether the resulting energy in the
center f mass frame $E_{c.m.}$ can grow unbounded, if we take into account
the processes in the white hole region. Before elucidating this issue, we
(i) describe basics of analysis of collision, (ii) possible scenarios of
collision that can be realized near the horizon Afterwards, we discuss,
which types of scenario correspond to the black hole region and which ones
are relevant for the white hole one.

The general analysis of collisions relies on the restrictions that come from
the conservation laws. We assume that such laws are valid in the point of
collision. This includes the conservation laws for the energy and angular
momentum:

\begin{equation}
E_{0}\equiv E_{1}+E_{2}=E_{3}+E_{4}\text{,}  \label{e12}
\end{equation}%
\begin{equation}
L_{0}=L_{1}+L_{2}=L_{3}+L_{4}.  \label{l12}
\end{equation}%
It follows from (\ref{e12}), (\ref{l12}) that%
\begin{equation}
X_{0}\equiv X_{1}+X_{2}=X_{3}+X_{4}.  \label{x12}
\end{equation}

There is also the conservation law for the radial momentum:%
\begin{equation}
\sigma _{1}P_{1}+\sigma _{2}P_{2}=\sigma _{3}P_{3}+\sigma _{4}P_{4}\text{.}
\label{r}
\end{equation}

We assume that particles 1 and 2 fall from infinity, so $\sigma _{1}=\sigma
_{2}=-1$. We are interested in high energy processes in which $E_{c.m.}$ is
unbounded since this is the necessary condition for $E$ to be unbounded as
well \cite{inf}, \cite{wald}. To this end, we choose particle 1 to be the
critical, particle 2 being usual since this gives rise to the unbounded $%
E_{c.m.}$ \cite{ban}, \cite{prd}. Then, one of particles (say, 3) is
near-critical and the \ other one (4) is usual \cite{up}, \cite{en}. All
possible scenarios can be described by two parameters - the sign of $C_{1}$
in (\ref{del}) and the value of $\sigma _{3}$ immediately after collision
(OUT for $\sigma _{3}=+1$ and IN for $\sigma _{3}=-1$). As a result, we have
4 scenarios OUT$+$, OUT$-$, IN$+$, IN$-$.

The first three were already analyzed in \cite{en}. In scenarios OUT$+$ and
OUT$-$ particle 3 after collision escapes immediately to infinity. In
scenario IN$+$ \ particle 3 continues to move inward after collision,
bounces back from the potential barrier and also returns to infinity. It
turned out \cite{up}, \cite{en} that the later scenario is especially
effective for the energy extraction. As far as IN$-$ is concerned, both
particles do not encounter a potential barrier and, therefore, fall into a
black hole. For this reason, scenario IN$-$ was rejected in \cite{en} since
no energy returns to infinity. However, now it is just this scenario which
we focus on. It corresponds to high energy propagation in the white hole
region (see below). Thus we have $\sigma _{3}=\sigma _{4}=-1$. We must
analyze the process under discussion for $N_{c}\rightarrow 0$ on the basic
of the conservation law (\ref{r}). In doing so, we follow the lines of Ref. 
\cite{en} applying the corresponding approach to the case that was not
considered there.

\section{Lower bounds on energy\label{low}}

If we collect the terms of the zeroth and first order in $N_{c}$ and take
into account the approximate expressions (\ref{cr}) - (\ref{ncr}), we obtain%
\begin{equation}
F=-\sqrt{E_{3}^{2}[(\frac{b}{h}-C_{1})^{2}-\frac{1}{h^{2}}]-m_{3}^{2}},
\label{F}
\end{equation}%
where%
\begin{equation}
F\equiv A+E_{3}(C_{1}-\frac{b}{h}),  \label{FA}
\end{equation}%
\begin{equation}
A=\frac{E_{1}b-\sqrt{E_{1}^{2}(b^{2}-1)-m_{1}^{2}h^{2}}}{h},  \label{A1}
\end{equation}%
\begin{equation}
C_{1}=\frac{b}{h}-\frac{A^{2}+m_{3}^{2}+\frac{E_{3}^{2}}{h^{2}}}{2E_{3}A},
\end{equation}%
\begin{equation}
F=\frac{A^{2}-m_{3}^{2}-\frac{E_{3}^{2}}{h^{2}}}{2A}.
\end{equation}

We are interested in scenario IN$-$. Then, $C_{1}<0$ gives us 
\begin{equation}
E_{3}^{2}-2E_{3}hA_{1}b+h^{2}(A^{2}+m_{3}^{2})>0,  \label{e2}
\end{equation}%
that can be rewritten as%
\begin{equation}
\left( E_{3}-\lambda _{+}\right) (E_{3}-\lambda _{+})>0,
\end{equation}%
\begin{equation}
\lambda _{\pm }=h[A_{1}b\pm \sqrt{A^{2}(b^{2}-1)-m_{3}^{2}}]\text{.}
\end{equation}

The condition $F<0$ gives us%
\begin{equation}
E_{3}^{2}>h^{2}(A^{2}-m_{3}^{2})\equiv \lambda _{0}^{2}\text{.}  \label{e0}
\end{equation}

If $\lambda _{\pm }$ are real, both bounds $E_{3}>\lambda _{+}$ and $%
E_{3}>\lambda _{0}$ are quite compatible with each other. If $\lambda _{\pm
} $ are complex, (\ref{e2}) and (\ref{e0}) are mutually consistent as well.
Thus there is no upper bound on $E_{3}$ and the SPP is possible.

As far as particle 4 is concerned, it has $E_{4}<0$. To obey the
forward-in-time condition (\ref{ftc}), it must have $L_{4}=-\left\vert
L_{4}\right\vert <0$. Then, $X_{4}=\left\vert L_{4}\right\vert \omega
-\left\vert E_{4}\right\vert $. Assuming that there is a flat infinity,
where $\omega \rightarrow 0$, we see that particle 4 either falls into
singularity or oscillates between turning points $r_{1}$ and $r_{2}$. In
doing so, it can intersect the horizons, thus appearing in new "universes"
due to a potentially rich space-time structure inside similarly to what
takes place for the Kerr metric \cite{ck}. However, under a rather weak and
reasonable restrictions on the properties of the metric, it cannot have more
than 1 turning point in the outer region, so the situation when $%
r_{+}<r_{1}\leq r\leq r_{2}$ is impossible. This was shown for the Kerr
metric in \cite{gpneg} and generalized in \cite{myneg}. For more information
about trajectories of particles 3 and 4, the metric should be specified.

The above treatment changes only slightly if we consider the Schnittman
process \cite{sch} when the critical particle 1 does not come from infinity
but moves from the horizon. Then, instead of (\ref{A1}), we should take $A=%
\frac{E_{1}b+\sqrt{E_{1}^{2}(b^{2}-1)-m_{1}^{2}h^{2}}}{h}$.

\section{Superenergetic particles}

In the above treatment, we tacitly assumed that all energies and angular
momenta are finite and do not grow unbounded when $N_{c}\rightarrow 0$. The
only place where $N_{c}$ appear in the relation between them are equalities (%
\ref{LE}), (\ref{del}), where it gives only small corrections. The above
approximate expressions for particle characteristics (\ref{cr}) - (\ref{ncr}%
) take into account this circumstances. In particular, for a usual particle, 
$X=O(1)$, the second term in the radical in (\ref{P}) has the order $%
N_{c}^{2}$. For a near-critical one, both terms in $P$ have the order $%
O(N_{c})$. Meanwhile, it turns out that there exists self-consistent
scenario, in which%
\begin{equation}
L_{3}=\frac{l_{3}}{\sqrt{N_{c}}}\text{,}  \label{Ll}
\end{equation}%
where $l_{3}$ is some coefficient not containing $N_{c}$. For small $N_{c}$, 
$L_{4}=L_{0}-L_{3}\approx -\frac{l_{3}}{\sqrt{N_{c}}}$.

It was found in \cite{center}, where it was pointed out that it corresponds
to falling both particles in a black hole, so it was put aside since we were
interested in particles returning to infinity. But now, it is this case that
came into play. Therefore, we take advantage of formulas already \ derived
in \cite{center} but exploit them in a new context - see eqs. (\ref{X3}), (%
\ref{be}) below. If (\ref{Ll}) is satisfied, the previous consideration
fails and the conservation law (\ref{r}) is to be analyzed anew. Now, for
particles 3 and 4 the second term in (\ref{P}) gives a small correction
(whereas for finite $L_{3}$ both terms for particle 3 would have the same
order), so%
\begin{equation}
P_{3,4}\approx \sqrt{X_{3,4}^{2}-N_{c}\frac{l_{3,4}^{2}}{\left( g_{\phi
}\right) _{H}}}\approx X_{3,4}-\frac{N_{c}}{2X_{3,4}}\frac{l_{3}^{2}}{\left(
g_{\phi }\right) _{H}}\text{. }
\end{equation}

Then, taking into account (\ref{cr}) - (\ref{us}) for particles 1 and 2, (%
\ref{e12}) - (\ref{x12}) and discarding the terms $O(N_{c}^{2}$) and higher,
one can show after algebraic manipulations that the following equation holds:%
\begin{equation}
\frac{l_{3}^{2}}{2\left( g_{\phi }\right) _{H}}(\frac{1}{X_{3}}+\frac{1}{%
X_{4}})=A\text{.}
\end{equation}%
Using again (\ref{x12}), one can obtain%
\begin{equation}
\left( X_{3,4}\right) _{c}\approx \frac{\left( X_{0}\right) _{c}}{2}(1\mp 
\sqrt{1-b})\text{,}  \label{X3}
\end{equation}%
where%
\begin{equation}
b\equiv \frac{2l_{3}^{2}}{\left( g_{\phi }\right) _{H}X_{0}A}\text{.}
\label{be}
\end{equation}%
It is implied that $b<1$.

It follows from definition (\ref{X}) that now 
\begin{equation}
E_{3}=\left( X_{3}\right) _{H}+\omega _{H}L_{3}\approx \left( X_{3}\right)
_{c}+\omega _{H}\frac{l_{3}}{\sqrt{N_{c}}}\text{.}
\end{equation}

Thus we have two usual particles which come down into a black hole. This is
contrasted with the standard case when particle 3 is near-critical and
returns to infinity.

We see that there are two energy scales for the SPP. On the first scale, $%
E_{3}$ can be as large as we like but with reservation that $E_{3}\ll \frac{%
const}{\sqrt{N_{c}}}$. On the second scale, $E_{3}\sim \frac{1}{\sqrt{N_{c}}}%
.$

In doing so, $E_{4}=E_{0}-E_{3}$ is negative having the same order $%
N_{c}^{-1/2}$. Discussion about the properties of the trajectories of such
particles from Section \ref{low} applies now as well.

\section{Conclusions}

Thus we showed that particle collision on our side of Universe (near the
black hole horizon) can lead to high energy fluxes on the other side. If,
vice verse, collision occurs in "another world", we can detect its
consequences in our one. We did not resort to the transformation between the
original stationary frame and the center of mass one. We would like to
stress that high energy behavior is found for the energies $E_{3}$ that can
be in principle detected in a laboratory. These results qualitatively agree
with claims made in \cite{ph} for the Kerr metric.

It is instructive to compare the situation for static charged and neutral
rotating black/white holes collecting the results of the present and
previous works \cite{rn}, \cite{whq}.

\begin{tabular}{|l|l|l|l|}
\hline
& SPP & $q_{3}$ & $L_{3}$ \\ \hline
charged black holes & yes & large & arbitrary \\ \hline
charged white holes & yes & large & $O(N_{c}^{-1/2})$ \\ \hline
neutral rotating black holes & no & $0$ & arbitrary \\ \hline
neutral rotating white holes & yes & $0$ & unbounded \\ \hline
\end{tabular}

Table 1. Conditions of the existence of the super-Penrose process.

We see that for the Reissner-Nordstr\"{o}m metric, if we compare collisions
near black and white holes, the situation is partially complementary to each
other. For finite $L_{i}$ (for example, with all $L_{i}=0$), there is the
SPP in the black \ hole case. However, it fails to exist near white holes.
As shown in \cite{whq}, only if particles with angular momenta $%
L_{3,4}=O(N_{c}^{-1/2})$ come into play, we obtain the SPP. Meanwhile, for
the rotating case, the SPP does not exist for black holes at all. Instead,
the white hole scenario opens new possibilities for the SPP, in which $E$
and $L$ of particles at infinity are unbounded. This happens on two scales:
on the first one $E$ and $L$ do not contain the parameter $N_{c}^{-1}$, on
the second scale they have the order $O(N_{c}^{-1/2})$ similarly to the
static charged case. The concrete properties of collisions are described by
somewhat different formulas and the type of energetic particles are
different: in the first case particle 3 is near-critical, in the second one
it is usual. However, in both cases the conserved energy $E_{3}$ is
unbounded.

The phenomenon under discussion is two-faced. On one hand, it shows that
high energy processes are indeed possible due to white holes and poses anew
the question about their potential role in nature. From the other hand, it
poses also a question about backreaction of such collisions on the metric
itself, including the fate of white holes.

\begin{acknowledgments}
This work was supported by the Russian Government Program of Competitive
Growth of Kazan Federal University.
\end{acknowledgments}

\end{document}